\documentclass[conference,hidelinks]{IEEEtran}
\IEEEoverridecommandlockouts
\usepackage{cite}
\usepackage{amsmath,amssymb,amsfonts}
\usepackage{algorithmic}
\usepackage{graphicx}
\usepackage{textcomp}
\usepackage{siunitx}

\ifCLASSOPTIONcompsoc
\usepackage[caption=false,font=normalsize,labelfont=sf,textfont=sf]{subfig}
\else
\usepackage[caption=false,font=footnotesize]{subfig}
\fi

\usepackage{tikz}
\usepackage{pgfplots}
\usepackage{pgfplotstable}
\usepackage{todonotes}
\pgfplotsset{compat=1.16} 

\usepackage{xcolor}
\definecolor{plot1}{HTML}{003f5c}
\definecolor{plot2}{HTML}{7a5195}
\definecolor{plot3}{HTML}{ef5675}
\definecolor{plot4}{HTML}{ffa600}

\definecolor{azure}{rgb}{0.0, 0.5, 1.0}
\definecolor{awesome}{rgb}{1.0, 0.13, 0.32}
\definecolor{asparagus}{rgb}{0.53, 0.66, 0.42}
\definecolor{cadetgrey}{rgb}{0.57, 0.64, 0.69}
\definecolor{amber}{rgb}{1.0, 0.49, 0.0}

\usepackage{hyperref}
\usepackage{listings}

\usepackage{booktabs}

\newcommand{\orcid}[1]{\href{https://orcid.org/#1}{\includegraphics[height=10pt]{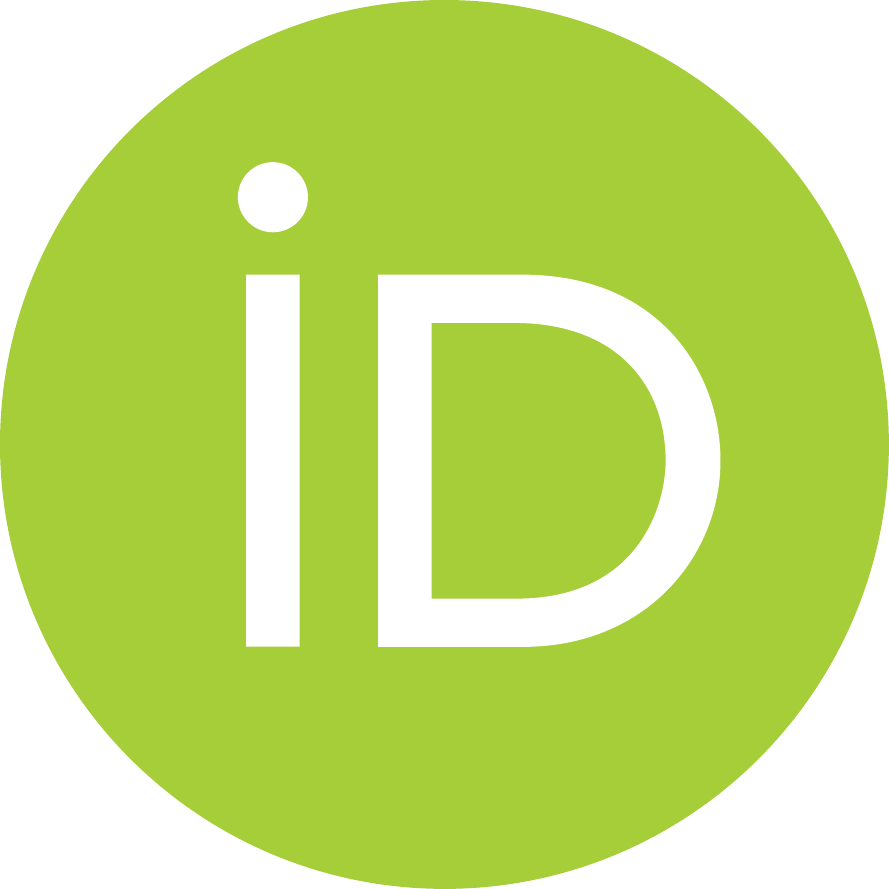}}}

\newcommand{\tm}{\textsuperscript{\texttrademark}}
\newcommand{\tr}{\textsuperscript{\textregistered}~}

\def\BibTeX{{\rm B\kern-.05em{\sc i\kern-.025em b}\kern-.08em
    T\kern-.1667em\lower.7ex\hbox{E}\kern-.125emX}}
\begin{document}
\bstctlcite{IEEEexample:BSTcontrol}

\title{Simulating Stellar Merger using HPX/Kokkos on A64FX on Supercomputer\ Fugaku}

\author{
\IEEEauthorblockN{Patrick Diehl\orcid{0000-0003-3922-8419}\IEEEauthorrefmark{1}\IEEEauthorrefmark{4}, Gregor Dai\ss\orcid{0000-0002-0989-5985}\IEEEauthorrefmark{2}, Kevin Huck\orcid{0000-0001-7064-8417}\IEEEauthorrefmark{3}, Dominic Marcello\IEEEauthorrefmark{1}, Sagiv Shiber\orcid{0000-0001-6107-0887}\IEEEauthorrefmark{4}, Hartmut Kaiser\orcid{0000-0002-8712-2806}\IEEEauthorrefmark{1},\\ and  Dirk Pfl\"uger\orcid{0000-0002-4360-0212}\IEEEauthorrefmark{2}}
\IEEEauthorblockA{
\IEEEauthorrefmark{1}LSU Center for Computation \& Technology, Louisiana State University,
Baton Rouge, LA, 70803 U.S.A. \\
Email: patrickdiehl@lsu.edu}
\IEEEauthorblockA{\IEEEauthorrefmark{2} IPVS, University of Stuttgart,
Stuttgart, 70174 Stuttgart, Germany}
\IEEEauthorblockA{\IEEEauthorrefmark{3} \textit{OACISS}, University of Oregon, Eugene, OR, U.S.A.}
\IEEEauthorblockA{\IEEEauthorrefmark{4} Department of Physics and Astronomy, Louisiana State University,
Baton Rouge, LA, 70803 U.S.A. }
}

\maketitle

\begin{abstract}
The increasing availability of machines relying on non-GPU architectures, such as ARM A64FX in high-performance computing, provides a set of interesting challenges to application developers. In addition to requiring code portability across different parallelization schemes, 
programs targeting these architectures have to be highly adaptable in terms of compute kernel sizes to accommodate different execution characteristics for various heterogeneous workloads. In this paper, we demonstrate an approach to code and performance portability that is based entirely on established standards in the industry. In addition to applying Kokkos as an abstraction over the execution of compute kernels on different heterogeneous execution environments, we show that the use of standard C++ constructs as exposed by the HPX runtime system enables superb portability in terms of code and performance based on the real-world Octo-Tiger astrophysics application. We report our experience with porting Octo-Tiger to the ARM A64FX architecture provided by Stony Brook's Ookami and Riken's Supercomputer\ Fugaku and compare the resulting performance with that achieved on well established GPU-oriented HPC machines such as ORNL's Summit, NERSC's Perlmutter and CSCS's Piz Daint systems. Octo-Tiger scaled well on Supercomputer\ Fugaku without any major code changes due to the abstraction levels provided by HPX and Kokkos. Adding vectorization support for ARM's SVE to Octo-Tiger was trivial thanks to using standard C++ interfaces.\end{abstract}

\begin{IEEEkeywords}
A64FX, HPX, AMT, Supercomputer\ Fugaku, Kokkos, SIMD, AMR, Stellar Merger
\end{IEEEkeywords}

\section{Introduction}

Over the last few years, the list of the top supercomputers has been increasingly dominated by GPU-based supercomputers. 
However, a notable exception is Riken's Supercomputer\ Fugaku. 
This ARM-based CPU-only supercomputer was the fastest machine on the Top 500 list and is still the number 1 machine on the HPCG Top\ 500 list\footnote{https://www.top500.org/lists/hpcg/2022/11/} as of November 2022. 

This presents interesting challenges for HPC developers, who have spent recent years adapting the codes more and more for large GPU clusters. There are several obvious yet important differences that need to be taken into account.
 Instead of having fewer but more powerful compute nodes containing several GPUs, we have a large number of comparatively weaker nodes, each containing a $48$-core ARM64FX CPU.
Instead of relying on CUDA\tm~(or CUDA-backends), compute kernels need to be adapted to make use of SIMD instructions.
Instead of having to use coarse-grained workloads for the GPU kernels, one can move to finer grained compute kernels suited for a few CPU cores (in turn providing opportunity for finer communication/computation interleaving and finer blocks of refinement in application using adaptive mesh refinement). 
And of course, lastly, instead of compiling for x86 architectures, HPC applications and their respective tool chain also need to work on ARM.

In this work, we focus on one such HPC application, Octo-Tiger, and on the changes required to run it on Fugaku, and we address the above challenges.
Octo-Tiger is an astrophysics code designed to model and study stellar mergers~\cite{marcello2021octo}. It does not rely on plain MPI, but instead is based on the task-based runtime system HPX~\cite{kaiser2020hpx}. 
Recent work on Octo-Tiger focused on the porting to GPU ranging from HPX-CUDA integrations~\cite{heller2016closing}, data-structure porting~\cite{Pfander18acceleratingFull}, HPX-Kokkos integrations~\cite{daiss2021beyond} and the porting of both the gravity- and the hydro solver to GPUs ~\cite{daiss2018octo, 10.1145/3295500.3356221, diehl2021octo}. 
Octo-Tiger has Kokkos compute kernels for all its major solvers to provide performance portability.

We have to compile and run HPX, Kokkos, and Octo-Tiger on Fugaku. There are two important pieces of previous work on this. We investigated automatic ways to adjust workload depending on whether we run on CPU or GPU by dynamically aggregating kernels for GPUs, simultaneously allowing fine-grained CPU kernels and thus finer adaptivity~\cite {daiss2022aggregation}.
We also already integrated ARM/SVE SIMD support into our Kokkos kernels using \lstinline[language=c++]{std::experimental::simd} and our own SVE types~\cite{arm_sve_simd}. However, we tested this only in node-level scenarios on the Ookami testbed, not in any distributed setting.

This builds the foundation for the present work, where we instead focus on the distributed case. In particular, we look at \textit{a)} the changes necessary to get everything running on Fugaku, \textit{b)} the impact of our SVE integration in distributed builds, \textit{c)} the impact of splitting our Kokkos compute kernels into more tasks (preventing starvation during tree traversals in the gravity solver), \textit{d)} the impact of communication optimizations regarding ghostlayers of neighbors on the same compute node.
Lastly, we take a look at recent, distributed runs of the same production scenarios on Fugaku, Piz Daint and Perlmutter.   

This work is best understood as a snapshot of the current 
developments regarding A64FX\tm~portability in (GPU-capable) HPX applications using Octo-Tiger as a representative example, showing and comparing its performance in systems based on A64FX and classical GPUs.
To our knowledge, this is both the first run with HPX on Supercomputer\ Fugaku, as well as the first application using Kokkos in distributed scenarios on this particular machine. 
Notably, getting HPX and Kokkos running on Supercomputer\ Fugaku was less effort than anticipated -- however, some adaptions were required to achieve better performance (such as making sure that the Kokkos kernels use explicit SIMD vectorization, allowing us to use the SVE SIMD types).

The remainder of this work is structured as follows:
In the next section, we start off with related work regarding Octo-Tiger.
In Section~\ref{sec:scientific:scenarios}, we will introduce the scientific application of Octo-Tiger, followed by our utilized software-stack in Section~\ref{sec:software:stack}.
In Section~\ref{sec:porting:a64fx}, we report on our experience to port Octo-Tiger to A64FX\tm. In Section~\ref{sec:distrbuted:scaling:all}, we show distributed results on Riken's Supercomputer\ Fugaku, NERSC's Perlmutter, and CSCS's Piz\ Daint. In Section~\ref{sec:ookami:performance:optimization}, we present potential performance optimizations for A64FX\tm. Finally, we follow up with a conclusion and future work on the remaining problems.


\section{Related work}

From the application's perspective, CASTRO~\cite{Almgren2020} is another astrophysics code with adaptive mesh refinement. Castro uses the MPI+X approach and relies on OpenMP for CPUs, CUDA\tm~for NVIDIA\tr GPUs, and HIP for AMD GPUs. However, we are not aware that Castro supports Kokkos and could not find scaling results on A64FX\tm. Sreepathi and et al.\ evaluated climate workloads on Supercomputer\ Fugaku, however, only node-level results were reported~\cite{9556014}. Furthermore, their focus is on climate models, while ours is on astrophysics. Thus we conclude that not many scientific codes used Kokkos on Supercomputer\ Fugaku. The other aspect of this work, is the C\texttt{++} Standard library for parallelism and concurrency (HPX). HPX is an asynchronous many-task system (AMT). We briefly highlight other AMTs with distributed capabilities, since the focus of this work is on distributed computing. Namely: Chapel~\cite{chamberlain2007parallel}, Charm\texttt{++}~\cite{kale1993charm++}, Legion~\cite{bauer2012legion}, Uintah~\cite{germain2000uintah}, and PaRSEC~\cite{bosilca2013parsec}. For a detailed comparison, we refer to~\cite{thoman2018taxonomy}. Charm\texttt{++} is the closest to HPX in terms of the implemented programming model. Overhead measurements using HPX and Charm\texttt{++} compared to MPI and OpenMP are shown in~\cite{https://doi.org/10.48550/arxiv.2207.12127}.

\section{Scientific scenarios}
\label{sec:scientific:scenarios}
In this paper, we study the parallel performance of Octo-Tiger for two major astrophysical scenarios in which we have used Octo-Tiger to simulate a stellar merger. In a stellar merger, two stars of a binary system interact, transfer mass, and finally merge to form a new single star with unique characteristics. In the first scenario, we simulate a system of a contact binary composed of two main-sequence (MS) stars. Its merger is assumed to produce a transient event like the well-documented one of V1309\ Sco (\emph{e.g.}, \cite{Tylenda2011}). In the second, we simulate a merger of a binary system consisting of two white dwarfs (double white dwarf system or DWD) believed to be the formation channel for a unique and rare type of stars, the R Coronae Borealis (RCB) stars (\emph{e.g.}, \cite{Crawford2020}, \cite{munson21}).
\subsection{V1309 Scorpii}
\label{sec:application:v1309}
Observational improvements in the last decade have driven further impetus to the study of binary systems by accurate and efficient numerical tools. Specifically, numerous transient observations, where their peak luminosities are below those of supernovae but still above those of novae, have called for the investigation of binary mergers.
In September 2008, for example, the contact binary, V1309\ Sco, produced a luminous transient attributed to a stellar merger. In this case, the brightness of the system increased by a factor of more than $10^4$ \cite{Tylenda2011}. Spectroscopic measurements classified this outburst as a luminous red nova \cite{Masonetal2010}. 
We have used Octo-Tiger to simulate the V1309 merger, both with gravity and hydrodynamics up until the merger. The orbital evolution is self-consistent, while the observational outburst can be deduced in post-processing. The simulations consist of tens of millions of grid cells.

The highest resolution model for the merger of V1309 has been achieved since then with an alternative approach, smooth particle hydrodynamics (SPH), with 170,000 particles \cite{Nandez2014}. 
In contrast, Octo-Tiger is based on adaptive mesh refinement. This allows one to resolve the atmosphere at a higher resolution, which will ultimately lead to better resolved temperature gradients at the surface of the stars and thus to a better light-curve estimation. The simulation of the V1309 merger in high resolution will provide us with greater insights into the nature of the mass and angular momentum flow and into the resulting circumstellar structure. 

\begin{figure}
    \centering
    \includegraphics[width=0.75\linewidth,trim={0 0 0 5cm},clip]{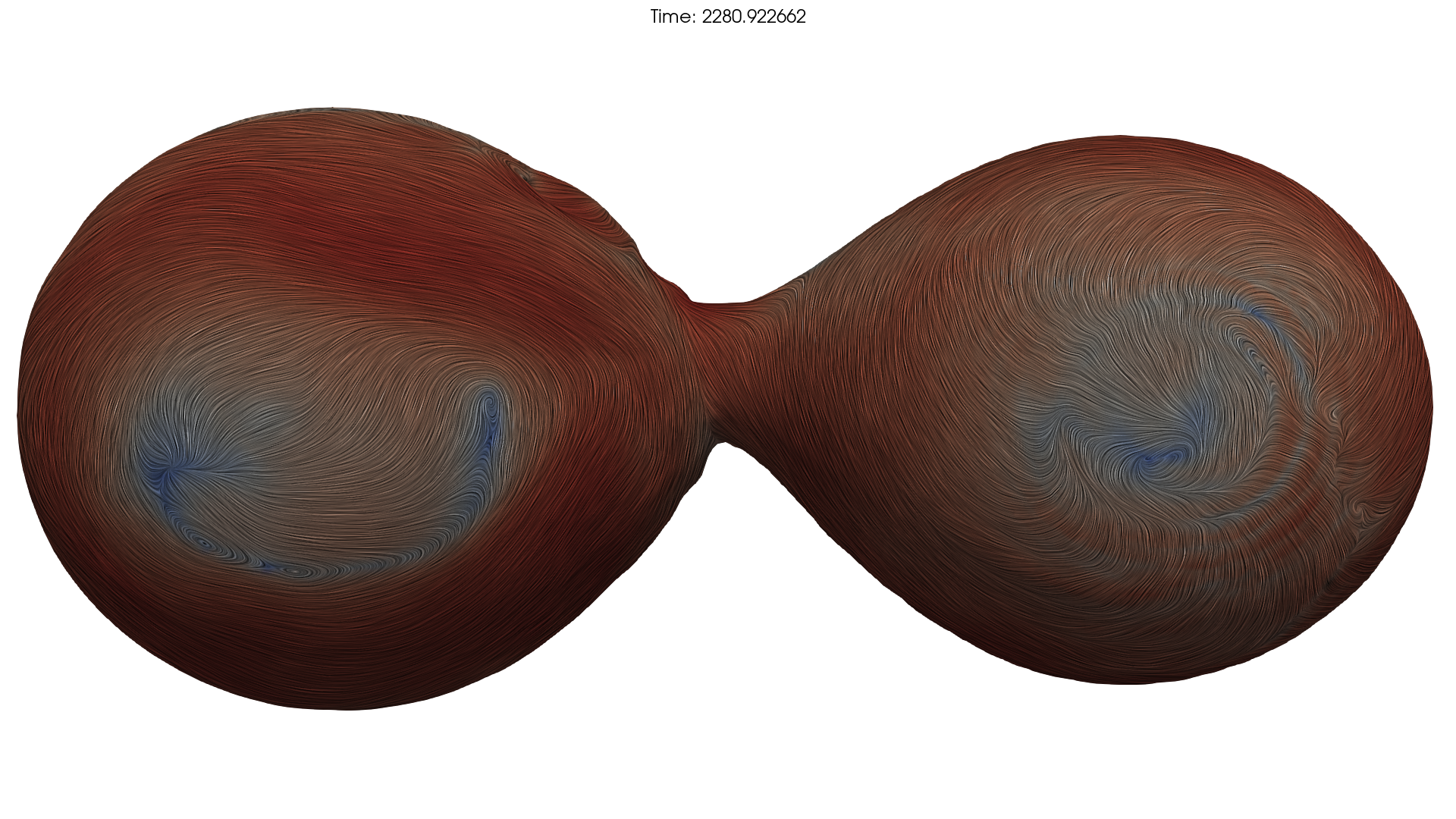}
    \caption{Double White Dwarf (DWD) system undergoing dynamical mass transfer. The surface is a density isosurface near the surface of the stars. The lines are streamlines. Color indicates velocity magnitude, with red being the largest and blue the smallest velocity. Adapted from~\cite{daiss2023sycl}.}
    \label{fig:isosurface}
\end{figure}

\subsection{Double White Dwarf Mergers (DWD)}
\label{sec:application:dwd}
RCB stars are intermediate-mass, hydrogen-deficient giants primarily made of helium. They are known to exhibit irregular and dramatic light variability in the form of deep declines that can leave the star at a minimum for several years before recovery is observed \cite{Clayton2012}. The two recent discoveries of very high Oxygen-18 to Oxygen-16 ratios, along with the relatively high average mass for these stars ($\sim0.9 M_{\rm \odot}$ \cite{Saio2008}), have elevated the merger of double white dwarfs (DWDs) as their most favorable formation scenario. We have used Octo-Tiger to simulate a number of DWD mergers in different resolutions and in which the mass ratio between the WDs is $0.7$, see Fig.~\ref{fig:isosurface} for an illustration. This mass ratio is not only appropriate for RCB stars, but by using the same mass ratio as previously, it allows a greater ability to compare our simulations with previous ones. We examine the resulted merger products, and estimate their probability to later evolve to a star with the characteristics of an RCB star. 

\section{Software stack}
\label{sec:software:stack}
\begin{figure}[tb]
    \centering
\begin{tikzpicture}
\node (Start) [draw] at (-2,0) {Octo-Tiger};

\node (silo) [draw] at (-5.5,-1) {\textcolor{azure}{Silo}};
\draw[->,thick] (Start) -- (silo);
\node (hdf5) [draw] at (-5.5,-2) {\textcolor{azure}{HDF5}};
\draw[->,thick] (silo) -- (hdf5);

\node (ccpuddle)[draw] at (-3.5,-1) {\textcolor{awesome}{cppuddle}};
\draw[->,thick] (Start) -- (ccpuddle);

\node (hpx) [draw] at (-2,-1)
{\textcolor{asparagus}{HPX}};
\draw[thick,->] (Start) -- (hpx);
\node (boost) [draw] at (-2,-2) {\textcolor{asparagus}{boost}};
\draw[->,thick] (hpx) -- (boost);
\node (hwloc) [draw] at (-3.5,-2) {\textcolor{asparagus}{hwloc}};
\draw[->,thick] (hpx) -- (hwloc);
\node (tcmalloc) [draw] at (-.5,-2) {\textcolor{asparagus}{tcmalloc}};
\draw[->,thick] (hpx) -- (tcmalloc);

\node (hk) [draw] at (0.25,-1) {\textcolor{awesome}{HPX/Kokkos}};
\draw[->,thick] (Start) -- (hk);
\draw[->,thick] (hpx) -- (hk);
\node (kokkos) [draw] at (1.5,-2) {\textcolor{awesome}{Kokkos}};
\draw[->,thick] (hk) -- (kokkos);

\end{tikzpicture}

\caption{Octo-Tiger's software stack and dependencies. In \textcolor{azure}{blue} the libraries Silo and HDF5 for IO. In \textcolor{asparagus}{green} HPX and its dependencies. In \textcolor{awesome}{red} HPX/Kokkos and cppuddle to support portability of performance on various CPUs and GPUs. For the specific versions, compilers, and build tools, we refer to Table~\ref{tab:compilers:libs}.}
\label{fig:software:stack}
\end{figure}
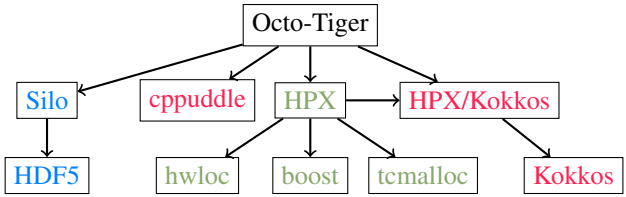

Octo-Tiger's software stack and dependencies are sketched in Figure~\ref{fig:software:stack}. The data structure (the octree) is saved to the hard disk using Silo's HDF file format. To ensure portability of performance, Octo-Tiger uses a combination of HPX and Kokkos. For the compute kernels on the CPU, Octo-Tiger uses Kokkos' HPX execution space. For the compute kernels on the GPU, Octo-Tiger uses the Kokko's CUDA\tm execution space to run on the NVIDIA\tr A100 on NERSC's perlmutter, on the NVIDIA\tr V100 on ORNL's Summit, and on the NVIDIA\tr P100 on CSCS's Piz Daint. Support for AMD GPUs using Kokkos is shown in~\cite{daiss2022aggregation} and Kokko's SYCL integration for Intel\tr GPUs is in development. Using HPX within Kokkos allows us to execute Kokkos compute kernels using asynchronous tasks. However, this does not allow us to asynchronously launch the Kokkos API and integrate the launches into HPX's asynchronous execution graph. Here, we use HPX-Kokkos to asynchronously call Kokkos functionality. All the components are described in more detail in the remaining part of this section. Table~\ref{tab:compilers:libs} shows all versions and compilers used for runs on the Supercomputer\ Fugaku and Ookami. For the versions used on Piz Daint and Perlmutter, we refer to~\cite{https://doi.org/10.48550/arxiv.2210.06437}.

\begin{table*}[tb]
    \centering
    \begin{tabular}{llllllllll}\toprule
    gcc     & 11.2.0/\textcolor{azure}{12.1.0} & hwloc &  1.11.12/\textcolor{azure}{2.8.0} & 
    boost   & 1.79.0/\textcolor{azure}{1.78.0} & Fujitsu MPI & 3.0/3.1   \\
    hdf5 & 1.8.12 & cmake & 3.19.5/\textcolor{azure}{3.24.2} & Vc & 1.4.1 & hpx & 1.7.1/1.8.1/\textcolor{azure}{\textit{b25e70b17c}}\\
    kokkos & \textit{2640cf70d}/\textcolor{azure}{\textit{7658a1136}} & hpx-kokkos & \textit{20a4496}/\textcolor{azure}{\textit{8ec88ae}} & SVE & a058275 & silo & 4.10.2 \\
    cppuddle & \textit{8ccd07a}/\textcolor{azure}{\textit{6e1715c}} & gperftools & \textit{bf8b714} & openmpi & \textcolor{azure}{4.1.4} & jemalloc & \textcolor{azure}{5.1.0} \\\bottomrule \\
    \end{tabular}

    \caption{Octo-Tiger (\textit{6848ea1}/\textit{8e42394}/\textcolor{azure}{\textit{1cfc36e9}}) was compiled with these compilers and libraries. The \textcolor{azure}{blue} version numbers show different versions used on Ookami.}
    \label{tab:compilers:libs}
\end{table*}

\subsection{C++ Standard library for parallelism and concurrency (HPX)}
As the asynchronous many-task system (AMT), we use the C\texttt{++} Standard Library for Parallelism and Concurrency (HPX)~\cite{kaiser2020hpx}. HPX is fully open source and available on GitHub\tr released under the boost software license\footnote{\url{https://github.com/STEllAR-GROUP/hpx}}. Comparing HPX to other AMTs, the major difference is that HPX's API is fully conforming with the recent C\texttt{++} 17/20 standard and provides experimental features from the upcoming C\texttt{++} 23 standard. Other AMTs are also implemented using C\texttt{++}, however, HPX's API is fully conformed to the C\texttt{++} standard for parallel algorithms and asynchronous programming. Some relevant features of HPX for this work are
\begin{itemize}
    \item HPX provides a uniform, standards-oriented API for ease of programming parallel and distributed applications.
    \item HPX provides a unified syntax for semantics for local and remote operations.
\end{itemize}
We use HPX as a backend for Kokkos to run HPX tasks on the lightweight threads, for example on A64FX. In addition, we use HPX/Kokkos to asynchronously launch Kokkos functionality and integrate within HPX's dependency graph.

\subsection{Kokkos and HPX-Kokkos}
\label{sec:hpx-kokkos}
Octo-Tiger uses Kokkos for portable compute kernels.
Kokkos is a C\texttt{++} framework that contains various abstractions to run compute kernels on various devices (CPU and GPUs) using different backends (Kokkos Execution/Memory Spaces)~\cite{9485033}.
With this framework, we can develop kernels that can run on all devices that we target, simply by writing the kernel once, but using different execution- and memory spaces depending on where we want to execute it.
To improve performance on CPU execution spaces, Kokkos also allows for explicit SIMD vectorization inside the kernels using C\texttt{++} types (which compile down to, for example, AVX512 instructions on CPU, and scalar instructions on GPU)~\cite{sahasrabudhe2019portable}. We showcased this for single node scenarios within Octo-Tiger using various SIMD types, both \lstinline{std::experimental::simd} types and Kokkos SIMD types, on different platforms in previous work~\cite{daiss2022simd}. This included SVE types on A64FX\tm.

We also chose Kokkos for Octo-Tiger because there are two HPX-Kokkos integrations that make the two software frameworks a good match.
On the one hand, an HPX execution space within Kokkos itself that runs a passed kernel not on OpenMP threads (like the Kokkos OpenMP execution space) or just on the current thread (such as the Kokkos Serial execution space), but on the HPX worker threads. Depending on the configuration passed at kernel launch, the kernel gets split into multiple tasks that might be processed by any HPX worker thread (or multiple tasks at the same time, depending on the scheduling and on the number of tasks available at the system currently).
On the other hand, there is also a second HPX-Kokkos integration~\cite{hpx_kokkos} that allows us to get HPX futures for any asynchronous launch of the Kokkos kernel, effectively allowing us to treat Kokkos kernels as HPX tasks (and thus define continuation tasks that will run once the original kernel completes).

Together, this effectively allows moving away from a fork-join style of launching compute kernels. Instead, during tree-traversals in Octo-Tiger any HPX task may asynchronously launch Kokkos kernels and define what should be done with the results by adding HPX continuations. The launched kernels may, in turn, run on any HPX worker thread as smaller HPX tasks, keeping the system saturated.
To give an idea about the amount of kernel launches per time-step: in Octo-Tiger, we usually have multiple ($>10$) kernel launches per sub-grid in each time-step. As we have thousands to millions of sub-grids (depending on whether we run a local test or full-scale production scenario), this means that we have numerous small kernels that are quickly launched. These fine-grained compute kernels can in turn be used to hide communication latencies.

In the context of this work this means that even though the Kokkos compute kernels were originally developed for GPUs, they can be used on A64FX\tm~ CPUs using SVE with no modifications (beyond choosing the SVE SIMD types at compile-time). Furthermore, for distributed scenarios where there is less and less work per compute node when the number of used compute nodes increases (especially during the tree-traversals in the gravity solver), the HPX execution space can be used to further split the compute kernels into smaller tasks to avoid all CPU cores on any given node are starved. Although using just one task per kernel launch can be beneficial for small kernels (since it runs the kernel on the current task that launched it, which usually benefits from a hot cache), splitting it into smaller tasks can show its benefits at scale when individual nodes are becoming starved.
\subsection{Octo-Tiger}


Octo-Tiger is a code for modeling self-gravitating astrophysical fluids. It uses a Eulerian grid-based approach to model the hydrodynamics and the fast multipole method (FMM) to compute the gravitational field. The grid structure for the hydrodynamics is based on an adaptive mesh refinement (AMR) octree, with each node being either a leaf node or a fully refined interior node of the octree. Each node is associated with an $N\times N \times N$ sub-grid containing the evolved hydrodynamic state variables ($N$ is typically $8$). These variables are evolved in time using a semi-discrete, finite volume scheme, coupled with a third order Runge-Kutta solver. To enable machine precision conservation of some of the evolved variables, Octo-Tiger does {\it not} use adaptive time stepping. We have additionally implemented features specifically suited to the study of interacting binary stars, such as rotating the AMR grid with the original orbital frequency of the binary. This reduces the numerical viscosity, at least in the early phases of a simulation. AMR is based on the density field and a field of tracer variables that track the original mass fractions of the binary components (\emph{e.g.}\ as the core and envelope fractions). Octo-Tiger can also refine the mesh on the basis of the density gradient.

The FFM part of the code piggybacks on the AMR structure of the hydrodynamics module, with $N\times N\times N$ sub-grids of multipoles, each colocated with an associated hydrodynamics sub-grid. The mass distribution of a hydrodynamics sub-grid cell is approximated as a single monopole containing all of its mass located at the center of the cell, while interior sub-grids contain monopole and quadrupole moments computed about the centers of mass of the cell. The original FMM conserves linear momentum to machine precision. We have added a modification that enables it to simultaneously conserve angular momentum. This requires Octo-Tiger to also compute the octupole moment with the lower moments. Although the hydrdynamics solver does \textbf{not} conserve angular momentum up to machine precision, the conservation of angular momentum enables gravity and hydrodynamic solvers to be coupled in a manner that conserves total energy (kinetic + internal + potential) up to machine precision. 

Our binary models are initialized using an iterative ``self-consistent field'' (SCF) technique. The hydrostatic equilbrium equation in the rotating frame is integrated to produce an algebraic equation with two unknowns, the ``effective'' gravitational potential and the enthalpy. The module is capable of producing detached, semi-detached, and contact binaries, such as the progenitor to V1309\ Sco. The structure of the components may be polytropic or a ``bi-polytropic'' structure, with core, envelope, and/or common envelope components.

Figure~\ref{fig:isosurface} depicts a model of a DWD undergoing dynamical mass transfer. The surface is a density isosurface near the surface of the stars. The lines are streamlines. Color indicates velocity magnitude, red being the largest and blue the smallest velocity.

\section{Porting the software stack to A64FX}
\label{sec:porting:a64fx}
Porting Octo-Tiger and its dependencies to A64FX\tm~was rather straightforward. First, HPX was tested on a small Raspberry Pi cluster~\cite{gupta2020deploying} and ported to the ARM architecture. Fugaku uses the Parallel Job Manager (PJM) for scheduling. HPX was extended to support PJM\footnote{\url{https://github.com/STEllAR-GROUP/hpx/pull/5870}}. The support for vectorization by Kokkos using \texttt{std::experimental::simd} was tested on the Ookami cluster 
with A64FX\tm~CPUs on a single node~\cite{https://doi.org/10.48550/arxiv.2210.06439} using the GNU compiler collection.

The recommended compilation process on Supercomputer\ Fugaku is to log into the x86 head nodes and to use cross compilation for Arm A64FX\tm~using the Fujitsu\tr compiler (FCC). Figure~\ref{fig:software:stack} shows Octo-Tiger's dependencies. One issue here was that not all dependencies support cross-compilation. For that scenario, the recommendation is to compile natively on the A64FX\tr~compute nodes using the GNU compiler collection. We compiled all dependencies using gcc and the Fusitju\tr MPI compiler wrappers provided by Rikken. Sreepathi et al. evaluated different compilers on A64FX\tm~and reported that gcc showed the best performance for their application~\cite{9556014}.

\section{Distributed scaling results}
\label{sec:distrbuted:scaling:all}
In this section, we look at the effect of Supercomputer\ Fugaku's boost mode on Oct-Tiger, see Section~\ref{sec:effect:boost}. After that, we first compare the distributed performance on Supercomputer\ Fugaku with the performance on CSCS's Piz Daint and ORNL's Summit using the v1309 scenario, see Section~\ref{sec:application:v1309}. Second, we compare the distributed performance on Supercomputer\ Fugaku with the performance on NERSC's Perlmutter using the DWD, see Section~\ref{sec:application:dwd}. Lastly, we look at distributed scaling on Supercomputer\ Fugaku up to $1024$ nodes, see Section~\ref{sec:distributed:scaling:fugaku}. For these runs, we enabled the SVE vectorization and the optimized communication on Supercomputer\ Fugaku. However, we could only implement these optimizations shortly before the end of the allocation period. We explore these optimizations in more detail on Ookami in Section~\ref{sec:ookami:performance:optimization}.   

\subsection{Effect of Supercomputer\ Fugaku's boost mode on Oct-Tiger}
\label{sec:effect:boost}
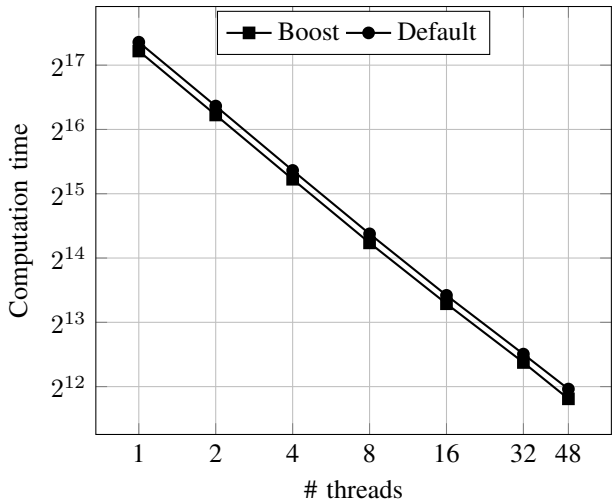
\begin{figure}[tb]
    \centering
    \begin{tikzpicture}
    \begin{axis}[xlabel={\# threads},ylabel={Computation time},grid,xmode=log,log basis x={2},xtick={1,2,4,8,16,32,48},xticklabels={1,2,4,8,16,32,48},ymode=log,log basis y={2},legend columns=2,legend style={at={(0.5,0.99)},anchor=north}]
    \addplot[thick,mark=square*,black] table [x expr=\thisrowno{0},y expr={\thisrowno{1}}, col sep=comma] {fugaku/fugaku-node-level.csv};
    \addplot[thick,mark=*,black] table [x expr=\thisrowno{0},y expr={\thisrowno{1}}, col sep=comma] {fugaku/fugaku-node-level-base.csv};
    \legend{Boost, Default};
    \end{axis}
    \end{tikzpicture}
    \caption{Node level scaling on a single Fugaku node. Fugaku has a power control function to save power on such a large system. The CPU frequency is reduced from  \num{2.2}\,\si{\giga\hertz} to \num{1.8}\,\si{\giga\hertz} in the default mode. However, there is a boost mode to use the full \num{2.2}\,\si{\giga\hertz}, but only for a limited amount of nodes. Octo-Tiger \textit{6848ea1} was used.}
    \label{fig:node:level}
\end{figure}

The CPU frequency of the A64FX\tm~ CPU is \num{2.2} \si{\giga\hertz}, however, on Supercomputer\ Fugaku the default CPU frequency is reduced to \num{1.8} \si{\giga\hertz}. For small node counts, there is a boost mode to use \num{2.2} \si{\giga\hertz}~\cite{kodama2020evaluation}. In Figure~\ref{fig:node:level} node-level scaling is performed at the node level for both modes. On a single node, the higher clock speed using the boost mode resulted in a marginal performance improvement. However, since we can not use the feature on higher node counts, for all remaining runs the default CPU clock speed was used. Table~\ref{tab:compilers:libs} shows the compilers and libraries used for all runs.

\subsection{Comparison with CSCS's Piz Daint and ORNL's Summit using the v1309 sceanrio}

Here, we compare the performance on Supercomputer\ Fugaku with previous results obtained on ORNL's Summit~\cite{diehl2021octo} and CSCS's Piz Daint~\cite{https://doi.org/10.48550/arxiv.2210.06437}. We ran the v1309 scenario, see Section~\ref{sec:application:v1309}, with $17$ million sub-grids. On Summit and Piz Daint, we used HPX's native CUDA\tm~ integration, since HPX/Kokkos was still during development during the compute time allocation period. The \textcolor{plot1}{blue} line shows the results on Summit using IBM\tr Power9\tm~ CPUs and NVIDIA\tr V100 GPUs. The \textcolor{plot2}{purple} line shows the results on NERSC's Perlmutter using NVIDIA\tr A100 GPUs. The \textcolor{plot3}{red} lines show the results on Supercomputer\ Fugaku using A64FX\tm~CPUs and no GPUs. Figure~\ref{fig:comparison:daint:summit:grids} shows the processed cells per second for all three suercomputers. On Summit, with $512$ GB memory per node, the complete spacerio is fitted into one node. On Piz Daint with $64$ GB memory per node, we could start with four nodes. On Supercomputer\ Fugaku with $28$ GB memory per node, we could start with $16$ nodes. On Summit with six NVIDIA\tr V100 per node, we obtained the best performance. The second best, on Piz Daint using one NVIDIA\tr P100 per node. However, the performance on Supercompuer Fugaku is close to the one on Piz Daint. On the other hand, this was expected due to the lower compute power of the A64FX\tm~CPU. Recall that we were running for the first time on A64FX\tm, and future code improvements might boost the performance. Figure~\ref{fig:comparison:daint:summit:speedup} shows the corresponding speed-ups with respect to the smallest number of nodes the scenario fits in.

\begin{figure}[tbp]
    \subfloat[\label{fig:comparison:daint:summit:grids}]{
    \begin{tikzpicture}
    \begin{axis}[xlabel={\# nodes},ylabel={Processed cells per second},title={v1309},grid,xmode=log,log basis x={2},xtick={1,2,4,8,16,32,64,128,256,512,1024,2048},ymode=log,log basis y={2},legend columns=2,legend style={at={(0.5,-.25)},anchor=north}]
    \addplot[thick,mark=*,plot1] table [x expr=\thisrowno{0},y expr={512*(35855+35904)/2*40/\thisrowno{1}}, col sep=comma] {summit/distributed-scaling-summit-no-apex.csv};
    \addplot[thick,mark=*,asparagus] table [x expr=\thisrowno{0},y expr={512*(35855+35904)/2*40/\thisrowno{1}}, col sep=comma] {pizdaint/distributed-scaling-no-apex-no-hyper.csv};
    \addplot[thick,mark=*,azure] table [x expr=\thisrowno{0},y expr={512*(35855+35904)/2*40/\thisrowno{1}}, col sep=comma] {fugaku/v1309.csv};
    \legend{Summit (CUDA), Piz Daint (CUDA), Fugaku (Kokkos), Perlmutter (Kokkos+Cuda};
    \end{axis}
    \end{tikzpicture}
    }
    \\
    \subfloat[\label{fig:comparison:daint:summit:speedup}]{
    \begin{tikzpicture}
    \begin{axis}[xlabel={\# nodes},ylabel={Speedup},title={Speedup (v1309)},grid,legend pos=north west,xmode=log,log basis x={2},xtick={1,2,4,8,16,32,64,128,256,512,1024,2048},ymode=log,log basis y={2},legend style={at={(0.5,-.25)},anchor=north},legend columns=2,ymin=1]
    \addplot[thick,mark=*,plot1] table [x expr=\thisrowno{0},y expr={1034.04/\thisrowno{1}}, col sep=comma] {summit/distributed-scaling-summit-no-apex.csv};
    \addplot[thick,mark=*,asparagus] table [x expr=\thisrowno{0},y expr={886.772/\thisrowno{1}}, col sep=comma] {pizdaint/distributed-scaling-no-apex-no-hyper.csv};
    \addplot[thick,mark=*,azure] table [x expr=\thisrowno{0},y expr={722.887/\thisrowno{1}}, col sep=comma] {fugaku/v1309.csv};
    \addplot[domain=1:128]{x};
    \addplot [domain=1:1800,dashed,shift={(axis cs:4,1)},legend image post style={shift={(0,0)}}]{x};
    \addplot [domain=1:1800,dotted,thick,shift={(axis cs:16,1)},legend image post style={shift={(0,0)}}]{x};
    \legend{Summit, Piz Daint, Fugaku,Ideal (Summit), Ideal (Piz Daint), Ideal (Fugaku)};
    \end{axis}
    \end{tikzpicture}
    }
    \caption{Comparison of the performance of ORNL's Summit, CSCS's Piz Daint, and Supercomputer\ Fugaku: \protect\subref{fig:comparison:daint:summit:grids} processed sub-grid  per second and \protect\subref{fig:comparison:daint:summit:speedup} the speedup $S$ with respect to the smallest number of nodes the scenario fitted in. Octo-Tiger \textit{584811a} was used.}
    \label{fig:comparison:daint:summit}
\end{figure}

\subsection{Comparison with NERSC's Perlmutter using DWD}
Figure~\ref{fig:performance:perlmutter} shows the processed cells per second for the DWD scenario, see Section~\ref{sec:application:dwd}, with a level of refinement of $12$ with \num{5150720} sub-grids. On NERSC's Perlmutter, Octo-Tiger was executed first using four NVIDIA\tr A100 and second using no GPUs. Here, we chose the level of refinement such that it fits into the $28$ GB of one Supercomputer\ Fugaku node. On Perlmutter, we used HPX's native CUDA\tm~support, since HPX Kokkos was still in development.
The \textcolor{plot1}{blue} lines show the results on Perlmutter and the \textcolor{plot2}{purple} line shows the results on Supercomputer\ Fugaku. Figure~\ref{fig:scaling:summit:beginning:12} shows the processed cells per second. The best performance yields the run on Perlmutter using all four GPUs per node. Not using the GPUs results in a drop of two orders of magnitude for the processed cells per second. Octo-Tiger clearly benefits from NVIDIA\tr A100 GPUs. The performance on Supercompuer\ Fugaku with this first attempt gets close to the CPU only run on Perlmutter. Future improvements may result in Fugaku's performance exceeding Perlmutter's.

Figure~\ref{fig:scaling:summit:close:12} shows the corresponding speedups with respect to the smallest number of nodes the scenario fitted in.

\begin{figure}[tbp]
    \centering
        \subfloat[\label{fig:scaling:summit:beginning:12}]{
    \begin{tikzpicture}
    \begin{axis}[xlabel={\# nodes},ylabel={Processed cells per second},title={DWD},grid,xmode=log,log basis x={2},xtick={1,2,4,8,16,32,64,128,256,512,1024,2048},ymode=log,log basis y={2},legend columns=2,legend style={at={(0.5,-.25)},anchor=north},xmax=128]
    \addplot[thick,mark=*,awesome] table [x expr=\thisrowno{0},y expr={10060*512/\thisrowno{1}/25}, col sep=comma] {perlmutter/level-12-gpu.csv};
    \addplot[thick,mark=diamond*,awesome] table [x expr=\thisrowno{0},y expr={10060*512/\thisrowno{1}/25}, col sep=comma] {perlmutter/level-12-cpu.csv};
    \addplot[thick,mark=square*,azure] table [x expr=\thisrowno{0},y expr={10060*512/\thisrowno{1}/25}, col sep=comma] {fugaku/level-12-dwd.csv};
    \legend{Perlmutter (CUDA), Perlmutter (CPU), Fugaku (Kokkos)};
    \end{axis}
    \end{tikzpicture}
    }
    \\
    \subfloat[\label{fig:scaling:summit:close:12}]{
    \begin{tikzpicture}
    \begin{axis}[xlabel={\# nodes},ylabel={Speedup $S$},title={Speedup (DWD)},grid,legend pos=north west,xmode=log,log basis x={2},xtick={1,2,4,8,16,32,64,128,256,512,1024,2048},ymode=log,log basis y={2},legend style={at={(0.5,-.25)},anchor=north},legend columns=2,ymin=1,xmax=128]
    \addplot[thick,mark=*,awesome] table [x expr=\thisrowno{0},y expr={133.438/\thisrowno{1}}, col sep=comma] {perlmutter/level-12-gpu.csv};
    \addplot[thick,mark=square*,awesome] table [x expr=\thisrowno{0},y expr={634.242/\thisrowno{1}}, col sep=comma] {perlmutter/level-12-cpu.csv};
    \addplot[thick,mark=*,azure] table [x expr=\thisrowno{0},y expr={1813.4/\thisrowno{1}}, col sep=comma] {fugaku/level-12-dwd.csv};
    \addplot[domain=1:256]{x};
    \legend{Perlmutter (CUDA),Perlmutter (CPU), Fugaku (Kokkos),Optimal};
    \end{axis}
    \end{tikzpicture}
    }
    \caption{Comparison of the performance of NERSC's Perlmutter and Fugaku: \protect\subref{fig:scaling:summit:beginning:12} processed sub-grid  per second and \protect\subref{fig:scaling:summit:close:12} the speedup $S$ with respect to a single node. Octo-Tiger \textit{584811a} was used.}
    \label{fig:performance:perlmutter}
\end{figure}

\subsection{Scaling on Fugaku}
\label{sec:distributed:scaling:fugaku}
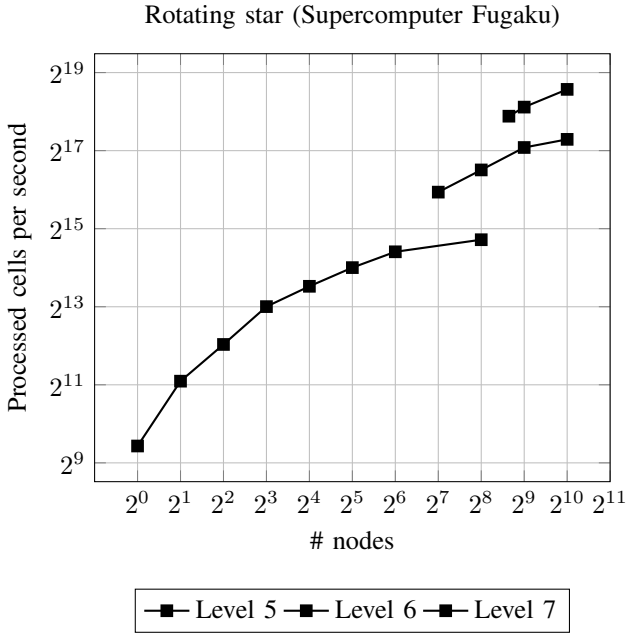
\begin{figure}[tbp]
    \begin{tikzpicture}
    \begin{axis}[xlabel={\# nodes},ylabel={Processed cells per second},title={Rotating star (Supercomputer\ Fugaku)},grid,xmode=log,log basis x={2},xtick={1,2,4,8,16,32,64,128,256,512,1024,2048},ymode=log,log basis y={2},legend columns=3,legend style={at={(0.5,-.25)},anchor=north}]
    \addplot[thick,mark=square*,black] table [x expr=\thisrowno{0},y expr={5048*512/\thisrowno{1}/10}, col sep=comma] {fugaku/star-5.csv};
    \addplot[thick,mark=square*,black] table [x expr=\thisrowno{0},y expr={27840*512/\thisrowno{1}/10}, col sep=comma] {fugaku/star-6.csv};
    \legend{Level 5,Level 6,Level 7};
     \addplot[thick,mark=square*,black] table [x expr=\thisrowno{0},y expr={172208*512/\thisrowno{1}/10}, col sep=comma] {fugaku/star-7.csv};
    \end{axis}
    \end{tikzpicture}
    \caption{Distributed scaling for the rotating star problem on Supercomputer\ Fugaku for level $5$ ($2.5$ million cells), refined to level $6$ ($14.2$ million cells), and level $7$ ($88.6$ million cells). These runs used some experimental support for SVE vectorization. In addition, we enabled the optimization for the communication. Octo-Tiger \textit{8e42394} was used.}
    \label{fig:fugaku:star:distributed}
   
\end{figure}

For the comparison with Perlmutter, we were limited to $128$ nodes during the test phase. Here, we show scaling on Supercomputer\ Fugaku up to $1024$ nodes. For these runs, we used the newly added SVE vectorization and enabled the optimization of the communication. However, we could get these features working on the last days of the test-bed allocation. We explore these optimizations in more detail on Ookami in Section~\ref{sec:ookami:performance:optimization}. We start with a rotating star with level $5$ ($2.5$ million cells), refined to level $6$ ($14.2$ million cells), and level $7$ ($88.6$ million cells) to scale up to $1024$ nodes. In addition, we used experimental support for SVE vectorization in the last days of the allocation. Figure~\ref{fig:fugaku:star:distributed} shows the processed cells per second for all levels. For level $5$, we start at one node and proceed to $256$ nodes. We start the level $6$ runs from $128$ nodes up to $1024$ nodes. We start the level $7$ run from $400$ nodes to $1024$ nodes. For level $6$, we see scaling up to $64$ nodes before we ran out of sufficient work per core. For level $6$, we scaled from $128$ nodes up to $512$ nodes before running out of sufficient work per core. For level $4$, we had enough work to scale up to $1024$ nodes. Table~\ref{tab:my_label} shows the average power consumption per node. However, Octo-Tiger started to hang for a larger node count. Due to the small test-bed allocation, we ran out of time to debug the hanging for larger node counts using Fujitsu\tr~MPI. 

\begin{table*}[tb]
    \centering
    \begin{tabular}{l|llllllllll} \toprule
     Level    & 2 & 4 & 8 & 16 & 32 & 64 & 128 & 256 & 512 & 1024  \\\midrule
     5    & 166.96 & 373.94 & 707.51 & 1145.69 & 1969.14 & 3933.05  & 11908.93 & 15228.07 & - & -  \\
     6 & - & - & - & - & - & - & 8659.86 & 19274 & 54261.58  & 111261.36 \\
     7 & - & - & - & - & - & - & - & - &   55310.55 & 111235.41  \\ \\
    \end{tabular}
    \caption{Average power consumption per node on Supercomputer\ Fugaku measured with PowerAPI~\cite{grant2016standardizing}}
    \label{tab:my_label}
\end{table*}



\section{Performance optimization on Ookami}
\label{sec:ookami:performance:optimization}
We tested the potential performance optimizations on Ookami, since the testbed allocation on Supercomputer\ Fugaku finished. However, we experienced rare deadlocks (in about $1$ out of $20$ runs) on distributed runs on Ookami which require further debugging. Note that we did not experience this on the same number of nodes on Supercomputer\ Fugaku. We use the rotating start with level $5$ again.

\subsection{Vectorization with SVE}
\label{sec:ookami:sve}
HPX was extended with support for vectorization using \lstinline{std::experimental::simd}~\cite{9651210}. Recently, Octo-Tiger was extended to use SIMD types within the Kokkos compute kernels, supporting \lstinline{std::experimental::simd} types, Kokkos SIMD and our own (\lstinline{std::experimental::simd}-compatible) SVE SIMD types~\cite{daiss2022simd}. Although still experimental, this allows us to use the SVE vectorization on A64FX\tm~for all major compute kernels in Octo-Tiger. 

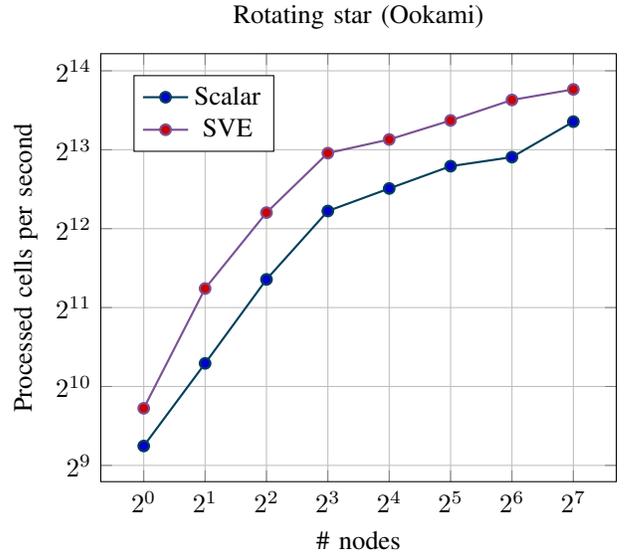
\begin{figure}[tbp]
    \centering
    \begin{tikzpicture}
    \begin{axis}[xlabel={\# nodes},grid,xmode=log,log basis x={2},xtick={1,2,4,8,16,32,64,128},ylabel={Processed cells per second},legend style={at={(0.2,0.95)},anchor=north},ymode=log,log basis y={2},title={Rotating star (Ookami)}]
    \addplot +[restrict expr to domain={\coordindex}{15:22},plot1,thick,mark=*]   table[x expr=\thisrowno{4}, y expr={5048*512/\thisrowno{5}/10}, col sep=comma]
    {ookami/ookami_optimization.csv};
    \addplot +[restrict expr to domain={\coordindex}{0:7},plot2,thick,mark=*]   table[x expr=\thisrowno{4}, y expr={5048*512/\thisrowno{5}/10}, col sep=comma]
    {ookami/ookami_optimization.csv};
    \legend{Scalar,SVE};
    \end{axis}
    \end{tikzpicture}
    \caption{Influence of vectorization using SVE on Ookami.}
    \label{fig:ookami:distributed:sve}
\end{figure}

As we use explicit SIMD vectorization with types, we can switch between scalar and SVE types at compile time, allowing us to easily judge the SIMD speedup by running the application twice: Once with the SVE SIMD types and once without.
On a single node, we could experience a speed-up between a factor of two and three for various parts of the code. In Figure~\ref{fig:ookami:distributed:sve}, we show the influence of vectorization on distributed runs up to $128$ nodes. We clearly see the effect of vectorization on the processed cells per second using the SVE vectorization, even though only the compute kernels are using it.

\subsection{Communication optimization}
\label{sec:ookami:communication}
\begin{figure}[tbp]
    \centering
    \begin{tikzpicture}
    \begin{axis}[xlabel={\# nodes},grid,xmode=log,log basis x={2},xtick={1,2,4,8,16,32,64,128},ymode=log,log basis y={2},ylabel={Processed cells per second},legend style={at={(0.2,0.95)},anchor=north},title={Rotating star (Ookami)}]
    \addplot +[restrict expr to domain={\coordindex}{8:14},thick,plot1,mark=*]   table[x expr=\thisrowno{4} , y expr={5048*512/\thisrowno{5}/10}, col sep=comma]
    {ookami/ookami_optimization.csv};
    \addplot +[restrict expr to domain={\coordindex}{0:7},thick,plot2,mark=*]   table[x expr=\thisrowno{4} , y expr={5048*512/\thisrowno{5}/10}, col sep=comma]
    {ookami/ookami_optimization.csv};
    \legend{OFF,ON};
    \end{axis}
    \end{tikzpicture}
    \caption{Influence of the optimization of the local communication for distributed runs.}
    \label{fig:ookami:distributed:communication}
\end{figure}
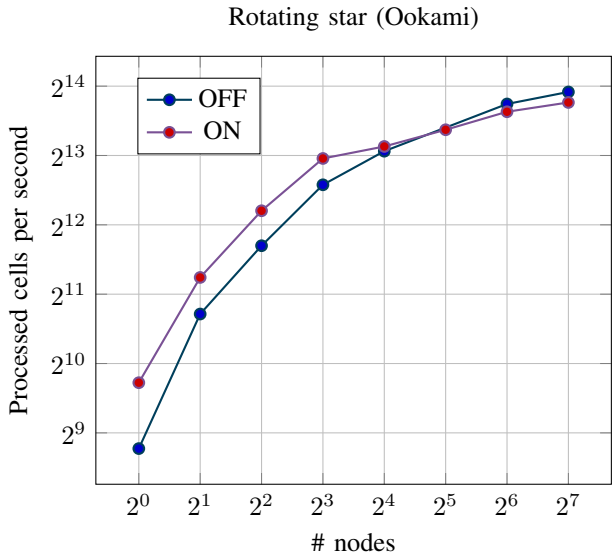

We further optimized the boundary communication in the hydro module as we detected a larger bottleneck here than expected during previous profiling runs. With this optimization, the sub-grids on the same HPX locality (process) are accessed directly in the memory, thus avoiding HPX actions and temporary communication buffers where possible. This reduces the total HPX action calls and the communication effort. However, some overhead of a different kind is added when collecting the hydro boundaries, as we need to make sure that the local neighbors are up-to-date. Here, we use simple local HPX promise/future pairs to notify neighbors when the local values are up-to-date and can be safely accessed. Figure~\ref{fig:ookami:distributed:communication} shows the processed cells per second with and without optimization enabled. For one, two, and four nodes, some benefit of communication is observed. On eight nodes, the break-even point is reached. After that, the optimization results in slightly worse performance.

\subsection{Usage of multiple HPX tasks per Kokkos kernel invocation}
\label{sec:ookami:splitting}
As mentioned in Section~\ref{sec:hpx-kokkos}, Kokkos includes an HPX execution space that allows splitting launched kernels into an arbitrary amount of HPX tasks.
Sometimes, it is preferable for small kernels to be launched as one HPX task (as this will run them on the current HPX worker thread, in turn making use of a potentially hot cache).
However, more compute-intensive kernels can, of course, benefit more from being split into multiple tasks. 
This is especially true when the amount of work per compute node is getting lower, and some CPU cores might simply run out of tasks to execute otherwise.

In Octo-Tiger, we use just one task per Kokkos kernel invocation by default, as each kernel works on just one sub-grid, which results in a low amount of work per individual kernel (but enables multicore usage by launching numerous kernels concurrently, each kernel working on a different sub-grid).
However, in Octo-Tiger's gravity solver, we can potentially benefit from splitting the Multipole kernels into more than one task in distributed scenarios, as it is possible to run out of work during tree-traversals here.
In each gravity solver iteration, we have one bottom-up tree traversal. In the second step, we then calculate the same-level cell-to-cell interactions on each tree level. Lastly, we do a third top-down step tree-traversal to compute the final results.
By splitting the calculation of the same-level cell-to-cell interactions (multipole interactions) on higher tree-levels into multiple task per sub-grid, we can avoid starvation of the CPU cores during these distributed tree-traversals.

\begin{figure}[tbp]
    \centering
    \begin{tikzpicture}
    \begin{axis}[xlabel={\# nodes},grid,,xmode=log,log basis x={2},xtick={1,2,4,8,16,32,64,128},ymode=log,log basis y={2},ylabel={Processed cells per second},legend style={at={(0.3,0.975)},anchor=north},title={Rotating star (Ookami)}]
    \addplot +[restrict expr to domain={\coordindex}{1:7},thick,plot1,mark=*]   table[x expr=\thisrowno{4} ,  y expr={5048*512/\thisrowno{5}/10}, col sep=comma]
    {ookami/ookami_optimization.csv};
    \addplot +[restrict expr to domain={\coordindex}{31:37},thick,plot2,mark=*]   table[x expr=\thisrowno{4} ,  y expr={5048*512/\thisrowno{5}/10}, col sep=comma]
    {ookami/ookami_optimization.csv};
    \legend{ON,OFF};
    \end{axis}
    \end{tikzpicture}
    \caption{Influence of Multipole work splitting using the Kokkos HPX execution space. \textit{OFF} means we just one HPX task per Multipole kernel, \textit{ON} means we use $16$ tasks per Multipole kernel to avoid starvation during the tree traversals.}
    \label{fig:ookami:distributed:sender}
\end{figure}
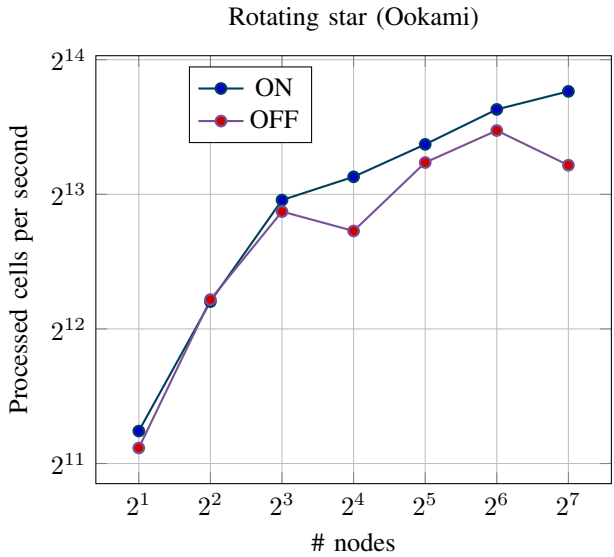

While using multiple tasks for one sub-grid (one kernel launch) is not yet beneficial when using just one compute node, it becomes important when we scale the same scenario to multiple nodes. As the amount of work per node decreases, being able to split compute kernels launches into multiple compute kernels can be of great benefit.
This can be observed in Figure~\ref{fig:ookami:distributed:sender}. Here we can see the difference between using just a single task for the Multipole kernel (same-level cell-to-cell interactions for refined sub-grids) launch and using 16 tasks per kernel launch instead.
On scenarios running on one compute node, using just one task per kernel invocation is sufficient, as the compute node has thousands of sub-grids to keep all cores busy.
However, this changes with 128 nodes, where using $16$ tasks per kernel invocation can yield a noticeable speedup over the other configuration, as it is easier to keep the cores of each compute node busy during the tree-traversals. This shows the benefit of the Kokkos HPX execution space, enabling us to choose how many tasks each kernel should be split into.

It should be noted that for these runs, we used an experimental version of the Kokkos HPX execution space, implementing a sender and receiver version\footnote{https://github.com/kokkos/kokkos/pull/5628}.

\subsection{Comparison between Supercomputer\ Fugaku and Ookami}
\begin{figure}[tbp]
    \centering
    \begin{tikzpicture}
    \begin{axis}[xlabel={\# nodes},grid,xmode=log,log basis x={2},xtick={1,2,4,8,16,32,64,128},ylabel={Processed cells per second},legend style={at={(0.7,0.4)},anchor=north},ymode=log,log basis y={2},title={Rotating star (Ookami \& Supercomputer\ Fugaku)}]
    \addplot +[restrict expr to domain={\coordindex}{15:22},plot1,thick,mark=*]   table[x expr=\thisrowno{4}, y expr={5048*512/\thisrowno{5}/10}, col sep=comma]
    {ookami/ookami_optimization.csv};
    \addplot +[restrict expr to domain={\coordindex}{0:7},plot1,thick,mark=square*]   table[x expr=\thisrowno{4}, y expr={5048*512/\thisrowno{5}/10}, col sep=comma]
    {ookami/ookami_optimization.csv};
    \addplot[thick,mark=*,plot2] table [x expr=\thisrowno{0},y expr={5048*512/\thisrowno{1}/10}, col sep=comma] {fugaku/star-5.csv};
    \legend{Ookami, Ookami (SVE), Fugaku};
    \end{axis}
    \end{tikzpicture}
    \caption{Comparison between Ookami and Supercomputer\ Fugaku}
    \label{fig:ookami:fugaku}
\end{figure}
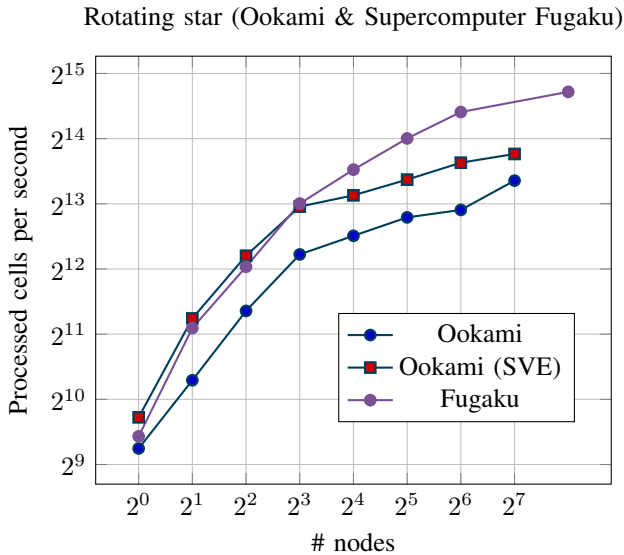

Figure~\ref{fig:ookami:fugaku} shows the comparison of the rotating star level $5$ on Ookami and Supercomputer\ Fugaku. The \textcolor{plot1}{blue} lines show the cells processed per second on Ookami. On Ookami, we used the fully optimized version with optimized communication, see Section~\ref{sec:ookami:communication}, and multipole work splitting, see Section~\ref{sec:ookami:splitting}. Here, we ran the code with and without SVE vectorization; see Section~\ref{sec:ookami:sve}. The \textcolor{plot2}{purple} line shows the cells processed per second on Supercomputer\ Fugaku using SVE vectorization. Here, we enabled communication improvement and an older version of SVE vectorization. The runs with the SVE vectorization are slighlty better on Ookami up to four nodes, since we optimzed the SVE vectorization after the Fugaku allocation ended. On 8 nodes the performance on both sytems is very close, however, after that the performance on Ookami is much better. One reason could be the additional enabled optimization for the multipole kernel. Furthermore, the Supercomputer Fugaku uses Fusitju\tr~Tofu-D interconnect and Ookami uses Infiniband interconnect. Here, further investigations are needed. However, we showcased some optimizations for A64FX CPUs.

\section{Conclusion}
In this paper, we explored two scenarios for stellar mergers using Octo-Tiger on Supercomputer\ Fugaku with A64FX CPUs using a Fugaku General Access Trial allocation. The goal of this work was to port Octo-Tiger using HPX and Kokkos to A64FX and obtain preliminary scaling results. Due to Kokkos' performance portability, there was no need to change Octo-Tiger's compute kernels. Porting Octo-Tiger and its software stack to A64FX required less effort than expected. However, some modifications were necessary for performance improvements. Note that Octo-Tiger was heavily used on x86 CPUs and NVIDIA GPUs for scaling and production runs before. The comparison with runs on NERSC's Perlmutter, ORNL's Summit, and CSCS's Piz Daint showed that Octo-Tiger benefits significantly from the usage of multiple GPUs on Summit ($6$ GPUs per node) and Perlmutter ($4$ GPus per node). However, the performance differences between Fugaku, Piz Daint ($1$ GPU per node) and Perlmutter without GPUs is not as dominant. Recall that Octo-Tiger was not optimized for A64FX, but rather for x86 CPUs and GPUs. It is notable that Octo-Tiger scaled on Supercomputer\ Fugaku without any major code changes due to the abstraction levels provided by HPX and Kokkos. We added the vectorization for SVE to Octo-Tiger and used the previous communication optimization to scale up to $1024$ Fugaku nodes. 

Furthermore, we investigated the effect of different optimizations on A64FX on Ookami. We show promising results of these optimizations. Finally, we compare the performance of Ookami and Supercomputer\ Fugaku using the same A64FX CPUs. However, Fugaku uses the Fujitsu\tr~Tofu-D interconnect with Fujitsu\tr~MPI and  Ookami uses Infiniband interconnect with OpenMPI.  

To conclude, due to the abstraction layers provided by HPX and Kokkos, we were able to easily port Octo-Tiger to A64FX on Supercomputer\ Fugaku. Depending on future access to Supercomputer\ Fugaku, we plan to obtain more results with our current optimizations. To further analyze the code performance, more runs using HPX's performance counters or Autonomous Performance Environment for Exascale (APEX)~\cite{huck2015autonomic} are needed.

\section*{Disclaimer}
{\footnotesize
The results on NERSC's Perlmutter were conducted on phase 1 such results should not reflect or imply that they are the final results of the system. Numerous upgrades will be made for Phase 2 that will substantially change the final size and network capabilities of Perlmutter. Furthermore, the runs were restricted to $128$ nodes. The results on Riken's Supercomputer\ Fugaku were conducted on a Fugaku General Access Trial allocation. The aim of this allocation was to get Octo-Tiger compiled, tested, and to do a preliminary scaling study within five months. 
}


\section*{Acknowledgment}
{\footnotesize
This research used resources of the National Energy Research Scientific Computing Center, the U.S. Department of Energy, Office of Science User Facility operated under Contract No. DE-AC02-05CH11231; This work used computational resources of the Supercomputer\ Fugaku provided by RIKEN through the HPCI System Research Project (Project ID: hp210311). This work was supported by a grant from the Swiss National
Supercomputing Centre (CSCS) under project ID s1078. The authors would like to thank Stony Brook Research Computing and Cyberinfrastructure, and the Institute for Advanced Computational Science at Stony Brook University for access to the innovative high-performance Ookami computing system, which was made possible by a \$5M National Science Foundation grant (\#1927880).}

\section*{Supplementary materials}
{\footnotesize
The Octo-Tiger build scripts are available on GitHub\footnote{\url{https://github.com/STEllAR-GROUP/OctoTigerBuildChain}}. The input files to run the v1309 scenario are available here\footnote{\url{https://doi.org/10.5281/zenodo.5213015}}. The DWD scenarios are available upon request, as an astrophysic paper is under preparation.}

\section*{Copyright notice}
\textcopyright 2023 IEEE. Personal use of this material is permitted.
Permission from IEEE must be obtained for all other uses, in any current or future  media, including reprinting/republishing this material for advertising or promotional purposes, creating new collective works, for resale or redistribution to servers or lists, or reuse of any copyrighted component of this work in other works. 

\bibliographystyle{IEEEtran}
\bibliography{references.bib}

\end{document}